\documentclass[aps,prd,twocolumn,superscriptaddress]{revtex4-1}
\usepackage[]{mathtools}
\usepackage{bm}
\usepackage{graphicx}
\usepackage{subfigure} 
\usepackage{float}
\usepackage[colorlinks,linkcolor=blue,anchorcolor=blue,citecolor=blue]{hyperref} 
\usepackage{verbatim}
\usepackage{booktabs}
\usepackage{multirow}
\usepackage{color}
\usepackage{mathtools}
\usepackage{longtable}
\usepackage{comment}
\usepackage{amssymb}
\usepackage{bm}
\usepackage{algorithm}
\usepackage{algpseudocode}
\usepackage{amsmath}

\renewcommand{\sec}[1]{\section{#1}}

\newcommand{\Msun}{\ensuremath{\mathrm{M}_\odot}}
\begin{document}


\title{Hierarchical Subtraction with Neural Density Estimators as a General Solution \\to Overlapping Gravitational Wave Signals}
\author{Qian Hu}
\email{Qian.Hu@glasgow.ac.uk}
\affiliation{Institute for Gravitational Research, School of Physics and Astronomy, University of Glasgow, Glasgow, G12 8QQ, United Kingdom}


\date{\today}

\begin{abstract}
  Overlapping gravitational wave (GW) signals are expected in the third-generation (3G) GW detectors, leading to one of the major challenges in GW data analysis. Inference of overlapping GW sources is complicated - it has been reported that hierarchical inference with signal subtraction may amplify errors, while joint estimation, though more accurate, is computationally expensive. 
  However, in this work, we show that hierarchical subtraction can achieve accurate results with a sufficient number of iterations, and on the other hand, neural density estimators, being able to generate posterior samples rapidly, make it possible to perform signal subtraction and inference repeatedly.   
  We further develop likelihood-based resampling to accelerate the convergence of the iterative subtraction. 
  Our method provides fast and accurate inference for overlapping GW signals and is highly adaptable to various source types and time separations, offering a potential general solution for overlapping GW signal analysis.
\end{abstract}
\maketitle

\sec{\label{sec0}Introduction}
The detection of gravitational waves (GWs) of nearly 100 compact binary coalescences (CBCs) by the LIGO-Virgo-KAGRA (LVK) collaboration~\cite{LIGOScientific:2016aoc, LIGOScientific:2017vwq, LIGOScientific:2018mvr, LIGOScientific:2018mvr, LIGOScientific:2020ibl, LIGOScientific:2021usb, KAGRA:2021vkt, LIGOScientific:2024elc} has contributed to the understanding of a variety of fundamental problems in physics and astrophysics, including the examinations of general relativity (GR)~\cite{LIGOScientific:2016lio, LIGOScientific:2018dkp, LIGOScientific:2019fpa, LIGOScientific:2020tif, LIGOScientific:2021sio}, matter in extreme conditions~\cite{LIGOScientific:2018hze, LIGOScientific:2018cki, LIGOScientific:2018ehx}, the astrophysical population of compact binaries~\cite{LIGOScientific:2018jsj, LIGOScientific:2020kqk, KAGRA:2021duu}, the measurement of cosmological parameters~\cite{LIGOScientific:2017adf, LIGOScientific:2021aug, DES:2019ccw, LIGOScientific:2019zcs}, the origin of $\gamma$-ray bursts~\cite{LIGOScientific:2017vwq, LIGOScientific:2017zic, ANTARES:2017bia}, {and so forth}. Many more GW detections are expected in the future: the third-generation (3G) detectors, such as the Einstein Telescope~\cite{Punturo:2010zz, Abac:2025saz} and Cosmic Explorer~\cite{reitze2019_CosmicExplorerContribution, Evans:2023euw}, will offer an order-of-magnitude improvement in sensitivity, enabling the detection of approximately $10^5$ CBCs per year with higher signal-to-noise ratios (SNRs) and longer durations~\cite{Branchesi:2023mws}. However, this scientific potential comes with challenges including increased computational demands~\cite{Hu:2024mvn}, the need to account for {the detecor's frequency-dependent response} (i.e., the Earth rotation~\cite{Baker:2025taj, Chen:2024kdc} and the finite detector sizes~\cite{Essick:2017wyl}), and systematic errors~\cite{Purrer:2019jcp,Hu:2022rjq, Hu:2022bji, Pizzati:2021apa, Himemoto:2021ukb, Samajdar:2021egv, Relton:2021cax, Relton:2022whr, Wang:2023ldq,Janquart:2022fzz}. 

Among these challenges, a particularly pressing issue in the 3G era is the presence of overlapping signals (OS)~\cite{Pizzati:2021apa, Himemoto:2021ukb, Samajdar:2021egv, Relton:2021cax, Relton:2022whr, Wang:2023ldq, Janquart:2022fzz}, which arises due to increased detection rates and extended signal durations. 
These overlaps can cause significant biases in parameter estimation (PE)~\cite{Pizzati:2021apa, Samajdar:2021egv, Wang:2023ldq} and downstream tasks such as testing GR~\cite{Hu:2022bji}. 
A number of studies have been conducted to understand the properties of OS. It is reported that OS's  {impacts on parameter estimation} could be significant when their merger time separation $\Delta t_c$ is less than $\sim1$\,s~\cite{Pizzati:2021apa, Samajdar:2021egv, Relton:2021cax, Wang:2023ldq} and when the OS have a mutual frequency band~\cite{Johnson:2024foj}, which depends on the GW source properties such as component masses and sky locations~\cite{Relton:2021cax, Gupta:2024lft, Wang:2023ldq}. 

Inference for OS can be approached through two main strategies: joint estimation and hierarchical subtraction~\cite{Janquart:2022fzz}. 
Following the widely-used Bayesian framework in GW astronomy~\cite{Veitch:2014wba}, the source parameters ${\theta}$ for the signal $h(\theta)$ is estimated by the samples drawn from the posterior probability distribution $p({\theta} | {d})$ given data ${d}$:
\begin{equation}
  \label{eq:posterior}
  p({\theta} | {d}, H) \propto p({d} | {\theta}, H) p({\theta}| H),
\end{equation}
where $p({\theta}|H) $ is the prior distribution {for parameters in the hypothetical model $H$}, $p({d} | {\theta}, H) $ is the likelihood function constructed by the idea that $d-h(\theta)$ is stationary Gaussian noise~\cite{Finn:1992wt}, i.e.,
\begin{equation}
  \label{eq:likelihood}
  p({d} | {\theta}, H) \propto \exp\left(-\frac{1}{2} (d-h(\theta)|d-h(\theta))\right) , 
\end{equation}
where $(a | b) = 4\mathrm{Re}\int_{0}^{\infty} a^*(f)b(f) / S_n(f) df$ is the noise-weighted inner product, with $S_n(f)$ being the one-sided power spectrum density (PSD) of the GW detector and $\mathrm{Re}$ denotes the real part. 
Assume two signals $A$ and $B$ are present in the data. Ideally, one should subtract both signals from the data in likelihood calculation, replacing the $h(\theta)$ in  Eq.~\ref{eq:likelihood} with $h(\theta_A)+h(\theta_B)$, effectively forming a joint PE for all parameters of the overlapping sources. However, this would result in a parameter space of over 30 dimensions, posing a considerable computational burden on stochastic sampling algorithms. 

On the other hand, hierarchical subtraction directly performs inference for the data with the single-signal likelihood Eq.~\ref{eq:likelihood} and obtains parameters of one of the sources (say, $\theta_A^{(0)}$), then subtracts $h(\theta_A^{(0)})$ from data and performs inference again, obtaining the second source $\theta_B^{(1)}$. This process can be repeated to generate $\theta_A^{(2)}$, $\theta_B^{(3)}$, \dots. Here superscript $(i)$ denotes results yielded from data with the $i$'th signal subtraction. However, this scheme faces several problems. First, it requires a series of PE runs which could be slow and expensive. Second, only the most-probable signal (i.e., a point estimate) is subtracted, which could be biased due to the statistical and systematic errors, leading to the error accumulation in hierarchical subtraction. In simulations performed by \citet{Janquart:2022fzz}, only 62\% of events showed improvements in $\theta_A^{(2)}$ compared with the initial estimate $\theta_A^{(0)}$. 

However, in this work, we show that the systematic errors in PE using hierarchical subtraction could diminish with sufficient iterations, and the practical problems in hierarchical subtraction can be overcome using a neural density estimator (NDE) trained for signal signals as the sampler.  NDEs rapidly generate posterior samples, enabling more iterations in hierarchical subtraction which helps the sampler approach the theoretical convergence. Moreover, the rapidness enables the subtraction of signals from data with multiple possible source parameters, generating an ensemble of signal-subtracted data and sampling for each of them, which  mitigates the inaccuracies in the subtracted signals. We also {use} a likelihood-based resampling scheme to further improve convergence.
Our method is built upon the well-established NDE-based PE for single signals~\cite{Dax:2021myb, Dax:2022pxd, Dax:2024mcn} and does not require expensive NDE training for different types of OS. It is adaptable to a wide range of source properties and time separations and therefore provides a fast and flexible solution for analyzing overlapping GW signals in the 3G era, which can be conveniently applied to all possible signal overlaps in the data, ensuring the robustness of PE against the impacts of OS. 
Details are given below.

\begin{figure}
  \includegraphics[width=0.48\textwidth]{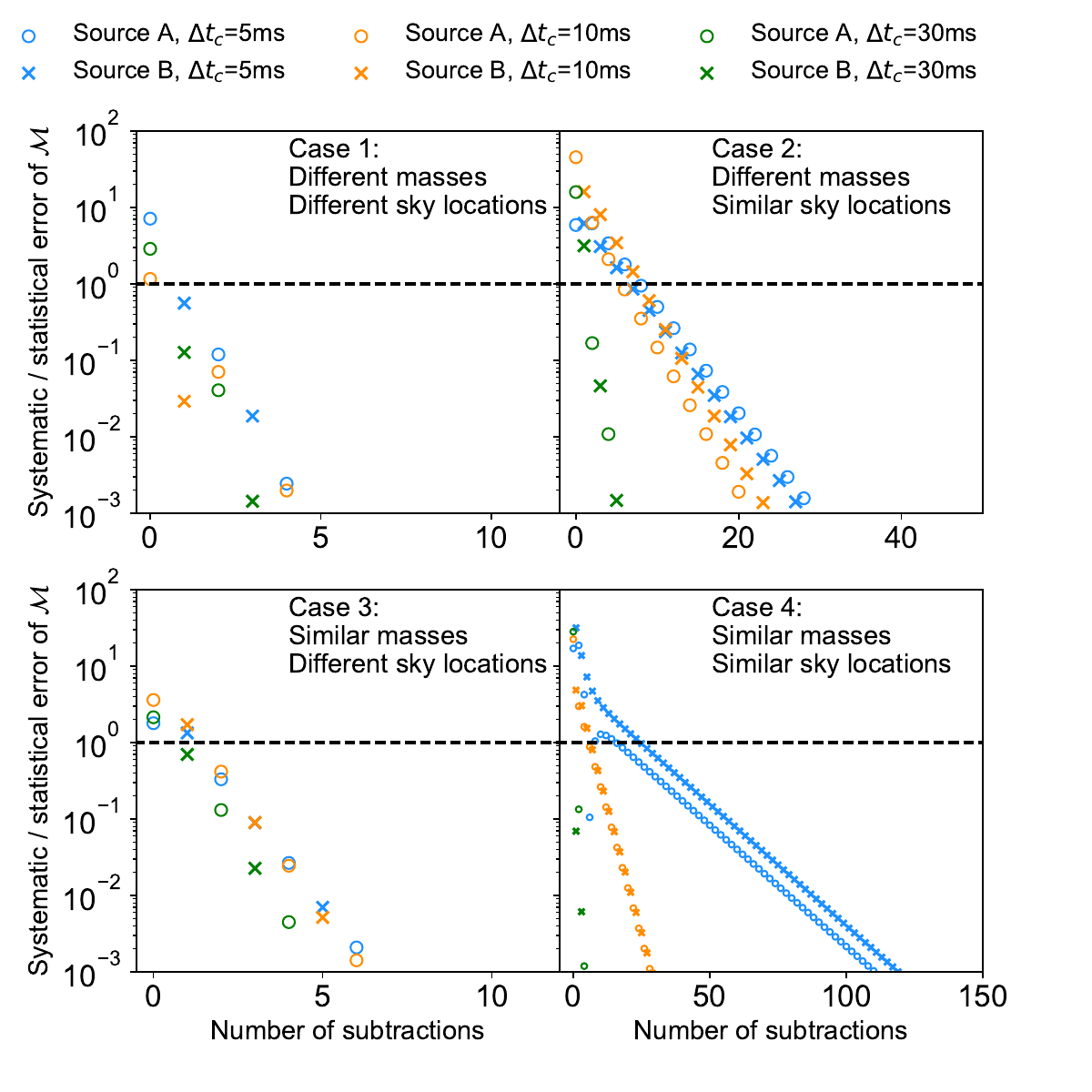}
  \caption{\label{fig:sys_evo} The ratio of systematic errors to statistical uncertainties in the chirp mass for various cases (as indicated by the figure labels) calculated by the Fisher matrix formalism
  A ratio greater than one indicates that the PE is dominated by systematic error. ``Similar'' refers to cases where the difference in parameters of two sources is $<5\%$, while ``different'' means $>50\%$. Injection parameter details are provided in Fig.\ref{fig:posterior_example}. Note that $\Delta t_c$ represents the arrivial time difference at the geocenter between two sources and is unrelated to Eq.\eqref{eq:syserror}.}
\end{figure}

\sec{\label{sec1}Systematic errors using hierarchical subtraction}
Systematic errors in PE arise from the residuals $\delta H$ after subtracting model signal $h_m$ (which does not equal to the true signal $h$) from data $d$, i.e. $d-h_m = n+\delta H$. 
As shown in Refs.~\cite{cutler2007_LISADetectionsMassive,antonelli2021_NoisyNeighboursInference}, the systematic error in the estimation of the $i$'th parameter using a maximum-likelihood estimator is given by
\begin{equation}
  \label{eq:syserror}
  \Delta \theta^i = (F^{-1})^{ij} (\partial_j h_\mathrm{m} | \delta H),
\end{equation}
where $F_{ij} = (\partial_i h|\partial_j h)$ is the Fisher matrix~\cite{cutler1994_GravitationalWavesMerging}, and repeated upper and lower indices denotes summation. $\partial_i$ denotes the partial derivative with respect to the $i$'th parameter. 
For OS, $\delta H$ corresponds to the other signal $h$ (if not subtracted) or the residual left after its subtraction, which can be approximated as $\Delta \theta^i  \partial_i h$ to the linear order. This framework enables an iterative calculation of systematic errors introduced by OS in hierarchical subtraction. Assuming $h_A$ is found first, we have: 
\begin{equation}
  \label{eq:sysevoA}
  \begin{aligned}
    &(\Delta \theta_A^i)^{(2m)}=\\
    &~~\begin{cases}
      (F_{A}^{-1})^{ij} (\partial_j h_A | h_B), & m=0\\
      (F_{A}^{-1})^{ij} (\partial_j h_A | \partial_k h_B (\Delta \theta_B^k)^{(2m-1)}), & m=1,2,3,\dots,
     \end{cases}
  \end{aligned}
\end{equation}
and 
\begin{equation}
  \label{eq:sysevoB}
  \begin{aligned}
    &(\Delta \theta_B^i)^{(2m+1)}= \\
    &~~(F_{B}^{-1})^{ij} (\partial_j h_B | \partial_k h_A (\Delta \theta_A^k)^{(2m)}), ~m=0,1,2,\dots,
  \end{aligned}
\end{equation}
where again superscripts in parentheses indicate the number of signal subtractions. We note that the statistical error can be averaged out so it is not taken into consideration. 
Using Eq.~\ref{eq:sysevoA} and \ref{eq:sysevoB}, we show the evolution of the ratio between systematic error and the statistical uncertainty $\sqrt{(F^{-1})^{ii}}$ during hierarchical subtraction in Fig.~\ref{fig:sys_evo}. It is shown that the systematic error vanishes with a sufficient number of iterations, even in the most challenging case where sources have similar masses and come from similar sky locations at nearly the same time. Such cases are rare: \citet{Wang:2025ckw} estimates that events with $\Delta t_c = 10$ ms occur less than once per year in the binary black hole (BBH) population. In most situations, systematic error becomes negligible after $\sim 5$ iterations. This means that hierarchical subtraction can be used to analyze OS in theory, while a practical implementation needs dedicated fast samplers, which, in this work, are neural density estimators. 

\sec{\label{sec2}Neural density estimators}
Neural networks can estimate probability densities~\cite{ruthotto2021introduction}. Specifically, conditional normalizing flows (CNFs)~\cite{kobyzev2020normalizing, papamakarios2021normalizing}, with their ability to provide exact likelihood, are often used in GW astronomy for inference tasks~\cite{Dax:2021myb, Dax:2021tsq, Dax:2022pxd, Dax:2024mcn,Gupte:2024jfe,Langendorff:2022fzq, Hu:2024lrj}. A CNF learns a conditioned bijective transformation between a simple base distribution (e.g., Gaussian) and a target distribution, which in our case is the GW posterior $p(\theta | d)$. Once trained, the CNF rapidly generates posterior samples by transforming draws from the Gaussian base distribution to the data space, yielding an approximate posterior $q(\theta | d)$. The discrepancy between  $q(\theta | d)$  and  $p(\theta | d)$  can be further reduced through importance sampling~\cite{Dax:2022pxd}. 

CNFs have demonstrated the ability to produce accurate and efficient posterior samples for real GW events~\cite{Dax:2021tsq, Dax:2022pxd, Dax:2024mcn}, including challenging scenarios such as inference with computationally expensive waveforms~\cite{Gupte:2024jfe} and long-duration signals~\cite{Dax:2024mcn, Hu:2024lrj}. For OS, \citet{Langendorff:2022fzq} proposed a CNF model to perform joint PE for two BBH signals arriving within 50 ms. 
However, the variety of potential overlapping scenarios — differing in arrival times, source types, priors, SNRs, and the number of signals — makes it less practical to train CNFs for every possible combination. This complexity is further increased by the high dimensionality of the parameter space and potential higher-order effects on the detector response~\cite{Baker:2025taj, Chen:2024kdc, Essick:2017wyl}.

As a proof of concept, we employ a publicly available CNF-based \texttt{DINGO} model~\cite{Dax:2021tsq}, which is trained on 8s non-overlapping BBH signals using the \texttt{IMRPhenomXPHM} waveform~\cite{Pratten:2020ceb} and the LIGO Hanford and LIGO Livingston detectors operating at the O1 sensitivity~\cite{abbott2019:GWTC1GravitationalWaveTransient}. We use samples generated by \texttt{DINGO} to perform hierarchical subtraction which is described in the next section. We note that the choice of NDE does not affect our algorithm as it is only used to estimate a single source assuming the other has been removed. Other NDEs, such as variational autoencoders~\cite{Gabbard:2019rde}, may also be applied to our algorithm. This guarantees flexibility to process various types of signals.

\sec{\label{sec3}Hierarchical subtraction with resampling}
Using NDEs trained for single signals, we can extract the first signal $\theta_A^{(0)}\sim q(\theta|d)$. Our goal is to alternately infer the posteriors of two overlapping signals while refining these estimates through hierarchical subtraction, i.e., $\theta_A^{(0)}$$ \rightarrow $$\theta_B^{(1)}$$\rightarrow $$\theta_A^{(2)}$$\rightarrow $$\theta_B^{(3)}$$\rightarrow\dots$. 
However, as discussed before, using a point estimate of the source parameters for signal subtraction may induce errors. We therefore randomly choose $N_{\mathrm{en}}=200$ parameters from the initial estimate, constructing an ensemble of possible source parameters. Each element in the ensemble will be used to subtract the signal, and the residual will be used to infer the parameters of the other source using the NDE. In other words, this process generates an evolving ensemble:
\begin{equation}
  \{\theta_A^{(0)}\} \rightarrow  \{\theta_B^{(1)}\} \rightarrow  \{\theta_A^{(2)}\} \rightarrow  \{\theta_B^{(3)}\} \rightarrow \dots
\end{equation}
Unless it is the first ensemble $\{\theta_A^{(0)}\}$, each element in the ensemble can contain multiple samples since NDEs are naturally parallelized to draw many samples at a time. In this case, we randomly choose one set of parameters from the element as the representative parameter, which is used for the next signal subtraction. Therefore, there are always $N_{\mathrm{en}}=200$ elements in the ensemble, with each element containing multiple samples but only one of the samples is used in hierarchical subtraction. At each iteration, all samples in each element are joined together as the PE result. We set the final number of samples to 16000, so each element in the ensemble contains 80 samples. 
This is not ideal - ideally, we should have $N_{\mathrm{en}}$ equal to the final number of samples and each element in the ensemble only contains one set of parameters, so that the one-to-one relationship between samples of two sources helps better capture the correlation between them. However, it takes $\sim 6$s for the \texttt{DINGO} model to draw samples (depending on the GPU model and NVIDIA A100 80GB is used in this work) and each iteration performs $N_{\mathrm{en}}$ \texttt{DINGO} sampling that currently cannot be parallelized on one GPU. With these limits, $N_{\mathrm{en}}=200$ already results in $\sim 20$ minutes per iteration, therefore a faster NDE model is required to achieve the ideal ensemble design and a faster speed. 

The initial estimate $\{\theta_A^{(0)}\}$ is often biased due to systematic errors caused by signals overlapping, the NDE sampling both signals or other issues of the NDE as it is not trained on such data. To mitigate this, we design a likelihood-based resampling for the first several iterations to remove the low-likelihood elements from the ensemble and replace them with high-likelihood ones. The likelihood of each element is calculated using their representative parameters:
\begin{equation}
  \label{eq:resampling_likelihood}
  \begin{aligned}
    &\{\log L^{(k)} \} =\\
    &~~\begin{cases}
      -\frac{1}{2}\| d-h(\{\theta_A^{(0)} \}) \| ^2, & k=0\\
      -\frac{1}{2} \| d-h(\{\theta_Y^{(k)} \}) -h(\{\theta_X^{(k-1)} \}) \|^2 , & k\geq 1,
     \end{cases} 
\end{aligned}
\end{equation}
where $\| a \| = \sqrt{(a | a)}$ and $(X,Y) = (A,B) ~\mathrm{or}~ (B, A)$: the signal being estimated in iteration $k$ is labelled as $Y$. Since the \texttt{DINGO} model marginalizes over the coalescence phase (i.e., each set of samples is 14-dimensional), we maximize the likelihood over uniform grids of 100 phase values to obtain an accurate estimate.
We can use a diagnostic statistic $\bar{L}^{(k)}=\mathrm{mean}\{ L^{(k)} \}$ to classify the number of signals and quantify the goodness of fit. We typically observe a leap of $\bar{L}^{(k)}$ at iteration 1 when the second signal is introduced to Eq.~\ref{eq:resampling_likelihood} and $\bar{L}^{(k)}$ would gradually increase during the hierarchical subtraction.
However, $N_\mathrm{en}$ points is a small set for the high-dimensional parameter space, so the $\bar{L}^{(k)}$ estimate is numerically unstable - it starts to fluctuate when the hierarchical subtraction is close to convergence and the new iteration brings little change to the posterior distribution. 
To ensure improving goodness of fit, we enforce a monotonic increase in $\bar{L}^{(k)}$ across iterations: since the representative parameter of each element in the ensemble is randomly chosen, we always repeat calculating the likelihood Eq.~\ref{eq:resampling_likelihood} until $\bar{L}^{(k)}$ surpasses that of the previous iteration. 

Then we perform resampling for elements in the ensemble $\{\theta_Y^{(k)} \}$ with weights based on tempered-likelihood $\{w \propto L^{1/T}\}$, where temperature $T$ is set to 10 in the first iteration and annealed to 1 exponentially. Each element (the representative sample along with other samples) in the ensemble is kept with the probability proportional to the weight, and this process is repeated until $N_\mathrm{en}$ elements are selected (there could be repeated elements in the ensemble). The temperature smoothens the likelihood function and effectively prevents resampling from being dominated by a small fraction of elements that can not represent the global posterior. The resampled ensemble will be used for signal subtraction to infer the parameters of the other source. We find the resampling technique greatly improves the stability and convergence of hierarchical subtraction. 

While resampling improves convergence, it may distort the parameter distribution because of the non-unit temperature. Moreover, when $\bar{L}^{(k)}$ fluctuates, it is not always possible to find higher-likelihood parameters within a reasonable number of attempts. If the number of retries exceeds a tolerance (set to 100), we stop using resampling and switch to a Metropolis-like sampling scheme. At each iteration  $k$ , for the signal  $Y$  being updated, we propose to accept new ensemble elements  $\{\theta_Y^{(k)}\}$  or retain old ones  $\{\theta_Y^{(k-2)}\} $ with acceptance probabilities: $\min \left(1, \{\frac{ L^{(k)}}{ L^{(k-2)}} \} \right)$. This acceptance rule ensures that $\bar{L}^{(k)}$ {are more likely to increase} over iterations.
Ideally, one can add a reverse jump term in the acceptance probability to form a Gibbs sampler that guarantees the convergence to the true joint distribution. However, this is unfeasible in our current framework because of the limits of the GNPE structure~\cite{Dax:2021myb} used in \texttt{DINGO} models: the flow likelihood is not directly accessible and sampling is relatively slow compared with a simple CNF. The hierarchical subtraction we presented in this work should be thought of as a greedy algorithm that approximates the joint parameter distribution, and the final result could be improved by importance sampling in principle. We leave these to our future study.

We terminate the hierarchical subtraction process when the Jensen-Shannon divergences (JSDs) between successive ensembles fall below a predefined threshold. Specifically, we monitor the JSD between  $\{\theta_Y^{(k)}\} $ and  $\{\theta_Y^{(k-2)}\}$ , as well as between  $\{\theta_X^{(k-1)}\}$  and  $\{\theta_X^{(k-3)}\}$. When both JSDs fall below  $10^{-4}$  nat, we consider the sampler to have converged.
Empirically, we find that most cases converge within 15 iterations, while challenging scenarios may require more than 20 iterations. A detailed pseudocode and an illustrative figure describing the full workflow are provided in the Appendix.

\begin{figure*}[h]
  \includegraphics[width=1\textwidth]{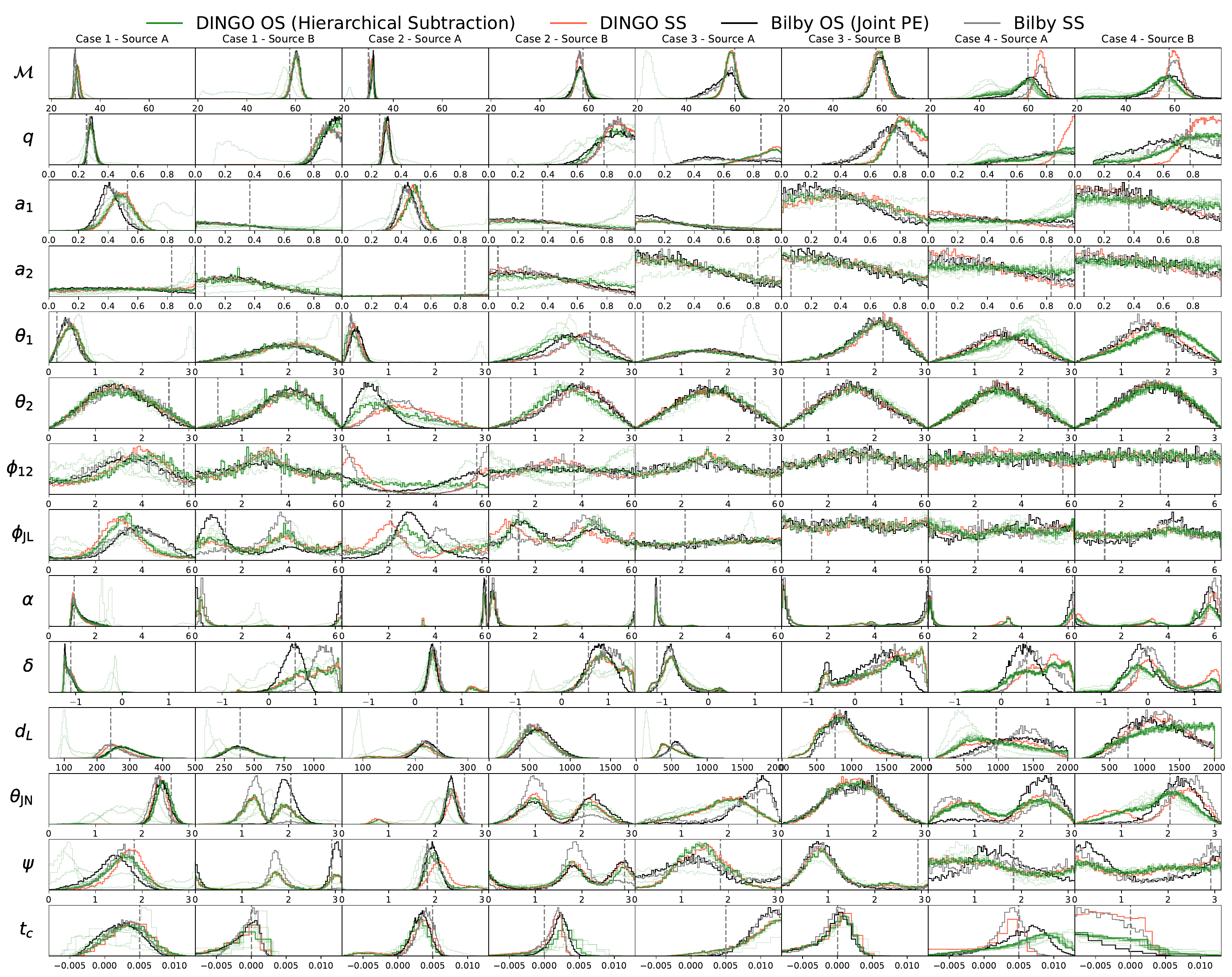}
  \caption{\label{fig:posterior_example} Posterior distributions of four cases considered in Fig.~\ref{fig:sys_evo} with the same injection parameters and $\Delta t_c = 5$. Green represents the posteriors obtained from the hierarchical subtraction for OS, while red represents the posteriors for isolated signals in the same noise realization using \texttt{DINGO}. Faint green lines show the hierarchical subtraction results during the first three iterations, illustrating early-stage biases. \texttt{Bilby} results for OS and SS are shown in black and grey, respectively. Grey dotted lines denote the true injection values.  }
\end{figure*}

\sec{\label{sec4}Results}
Using the hierarchical subtraction scheme described above, we analyze the four cases presented in Fig.~\ref{fig:sys_evo} using the same injection parameters in the $\Delta t_c = 5$ ms scenario, which represents the most challenging scenario in Fig.~\ref{fig:sys_evo}. The resulting posteriors are shown in Fig.~\ref{fig:posterior_example}. To assess whether the OS is efficiently removed, we also consider the single signal (SS) scenario where the other signal is perfectly removed from the original OS data. Both OS and SS data are analyzed using \texttt{DINGO} and Bayesian inference package \texttt{bilby}~\cite{Ashton:2018jfp} with the \texttt{dynesty} sampler~\cite{Speagle_2020} and the same prior used in the \texttt{DINGO} O1 model. Posterior distributions from \texttt{bilby} are considered the ground truth for OS and SS analyses. With one CPU and one GPU, the hierarchical subtraction OS analysis converges at 9-16 iterations in our simulations, corresponding to a total inference time of several hours, while the full joint PE runs take more than 20000 CPU core hours. 

In Fig.~\ref{fig:posterior_example}, the difference between \texttt{bilby} OS and \texttt{bilby} SS results shows how much impacts the overlapping source acts, with case 4 being the most impacted case.  The difference between \texttt{DINGO} OS and \texttt{DINGO} SS shows to what extent the overlapping signal is removed using our hierarchical subtraction scheme. In cases when the overlapping sources are not strongly correlated (e.g. Case 3), \texttt{DINGO} OS gives similar results as \texttt{DINGO} SS, implying that the hierarchical subtraction scheme efficiently removed the OS. In the contrary situation, \texttt{DINGO} OS starts to deviate from \texttt{DINGO} SS and converge to \texttt{bilby} OS (e.g. Case 4), suggesting that the correlation between sources is captured during hierarchical subtraction. 

While hierarchical subtraction successfully resolves some key parameters such as chirp masses and arrival times and can distinguish the two sources, some estimates are not always aligned with the ground truth \texttt{bilby} OS. This is because \texttt{DINGO} model does not provide perfectly accurate posterior samples, which can be seen from the difference between \texttt{bilby} SS and \texttt{DINGO} SS results. In principle, the samples generated by \texttt{DINGO} (both OS and SS) can be reweighted by importance sampling~\cite{Dax:2022pxd}, which greatly improves their accuracy. We leave this to our future exploration. 

Fig.~\ref{fig:posterior_example} also shows the consequence of neglecting OS effects when there are other signals present: the \texttt{DINGO} OS estimates are biased in the first several iterations. For instance, the luminosity distance is often underestimated due to the amplified signal from the OS, and sky directions could be inaccurate when overlapping sources are from different sky positions due to the messed-up timings in different detectors. Spin magnitudes tend to be overestimated because of the similarity between OS interference and spin precessing. 
These errors systematically diminish with subsequent iterations, and the sampler eventually converges to the \texttt{DINGO} SS result, \texttt{bilby} OS result, or somewhere in between. 


\begin{figure}
  \includegraphics[width=0.48\textwidth]{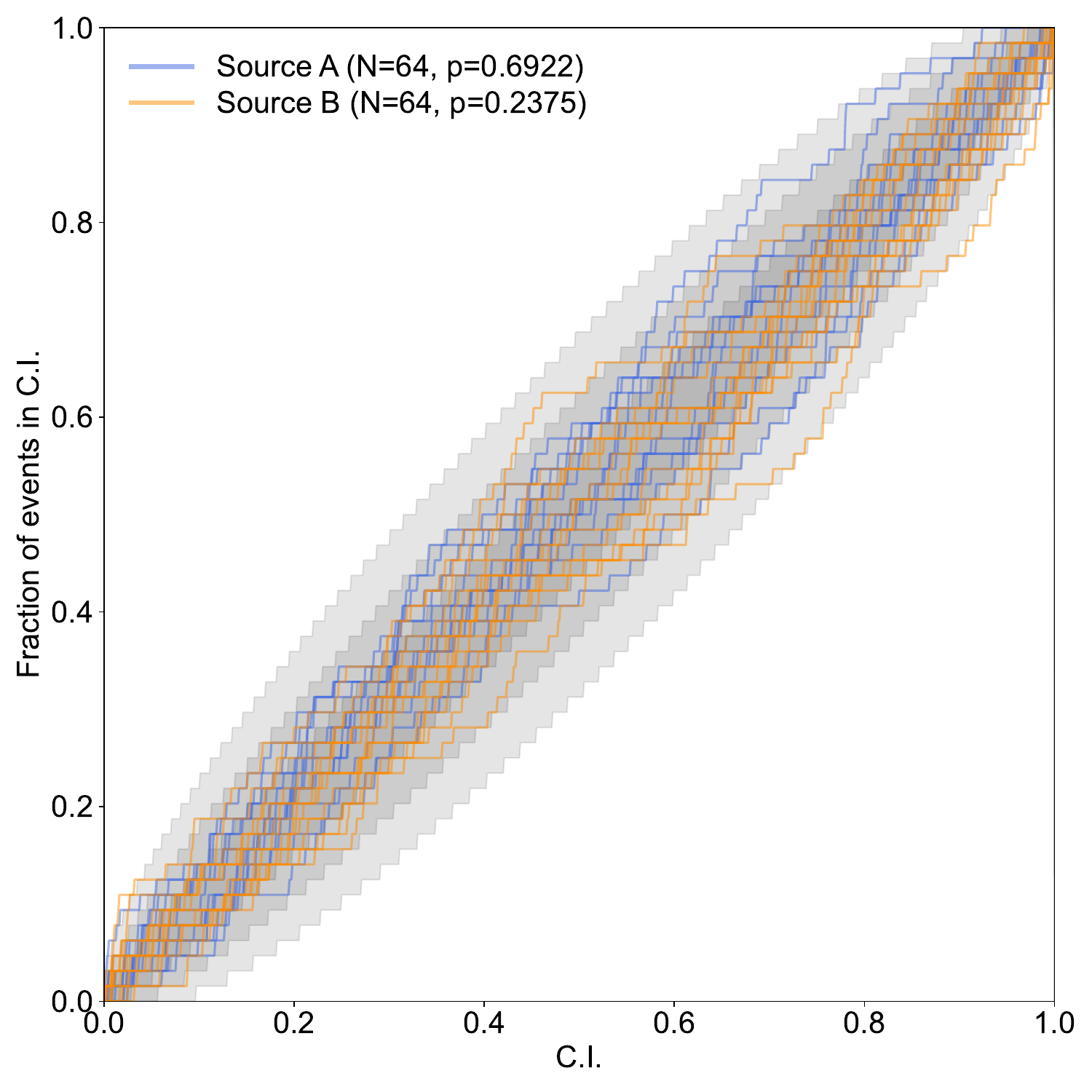}
  \caption{\label{fig:pp_plot} P-P plot for the hierarchical subtraction scheme applied to overlapping BBH signals. Each line represents a different parameter, with source A and B distinguished by color. The $x$-axis represents the credible interval (CI) and the $y$-axis indicates the fraction of simulations where the true value falls within the corresponding CI. A total of 64 BBH pairs are analyzed, with time separations  $\Delta t_c$  ranging from 5 to 50 ms. $p$-values in the legend are obtained from Kolmogorov-Smirnov tests between true parameter's confidence level in the posterior and a uniform distribution. $p<0.05$ represents a failure of P-P test.}
\end{figure}

To assess the self-consistency of our algorithm, we perform simulations of 64 BBH pairs within the training prior of the \texttt{DINGO} model, in which component masses range from 10 to 80 \Msun, luminosity distances extend to 2000 Mpc. Full spin precession and higher modes are included. We consider 4 time separations between coalescences:  $\Delta t_c = 5, 10, 30$, and 50 ms, with 16 BBH pairs simulated for each value. We set network SNR $>12$ as the detection criterion. 
The P-P plot is shown in Fig.~\ref{fig:pp_plot}, confirming our algorithm is self-consistent for OS.

\sec{\label{seclast}Conclusions and discussions}
We have demonstrated that PE for overlapping GW signals can be solved hierarchically with fast PE models (such as NDEs) trained on \textit{single} signals. 
Although hierarchical subtraction is traditionally considered less accurate than full joint PE, NDEs allow the process to iterate rapidly and enable the subtraction of an ensemble of possible signals, effectively overcoming the problems in hierarchical subtraction. Since NDEs are not explicitly trained on overlapping sources, they may exhibit issues such as degraded sampling efficiency or confusion between sources. We address this with a likelihood-based resampling technique and Metropolis updates in this work.
As shown in Fig.~\ref{fig:posterior_example}, our algorithm successfully recovers tightly overlapping sources from similar sky directions with only 5ms separation at the geocenter, matching theoretical expectations (Fig.~\ref{fig:sys_evo}). 

With $10^5-10^6$ CBC sources being detected annually in the 3G era, the computational cost of PE using traditional stochastic samplers becomes prohibitive~\cite{Hu:2024mvn}. 
Furthermore, $\gtrsim 10\%$ of BBHs may experience overlapping signals within $\Delta t_c < 4$s~\cite{Hu:2022bji}, posing additional challenges for both computational cost and accuracy. 
While some overlaps, such as those with $\Delta t_c > 1$s or NSBH–BNS combinations, are typically assumed to have small impacts, rare scenarios like loud signals masking faint ones still require validation. 
Our method, built on well-established PE models for isolated signals, offers a scalable and practical solution. As machine learning tools for GW PE continue to mature, fast posterior samplers are expected to be available for a wide range of sources in the 3G era. This makes it feasible to apply our method universally, ensuring that the impact of OS is accounted for in every source, without retraining models (which costs several weeks) or resorting to full joint PE (which can demand tens of thousands of CPU hours). These make comprehensive PE in the 3G era computationally feasible while preserving scientific accuracy.


Several practical limitations remain and there are potential improvements for our hierarchical subtraction algorithm. Currently, the number of signal realizations used for subtraction is limited to 200, which is significantly fewer than the total posterior samples, breaking the one-to-one correspondence in hierarchical subtraction and complicating the tracking of inter-source correlations. This could be resolved by using fast and parallelizable CNF PE models (e.g.,~\cite{Dax:2024mcn, Hu:2024lrj}) to enforce exact subtraction for every sample. CNFs also provide explicit likelihoods of each sample, simplifying the implementation of importance sampling which improves the accuracy of our OS PE. In addition, further investigation is required in particularly rare or challenging situations such as cases with $\Delta t_c < 5$ ms or more than two signals present. As this work serves as a proof of concept, addressing such edge cases is left to future studies.

\begin{acknowledgments}
The author would like to thank John Veitch and Michael Puerrer for helpful discussions and suggestions. The author is grateful for GPU resources provided by the LIGO Lab at Caltech which is supported by National Science Foundation Grants PHY-0757058 and PHY-0823459 and for CPU resources provided by Cardiff University funded by STFC grant ST/I006285/1 and ST/V005618/1. QH is supported by STFC grant ST/Y004256/1. 
\end{acknowledgments}


\appendix
\section{Hierarchical subtraction workflow \label{apdxa}}

\begin{figure*}
  \includegraphics[width=0.95\textwidth]{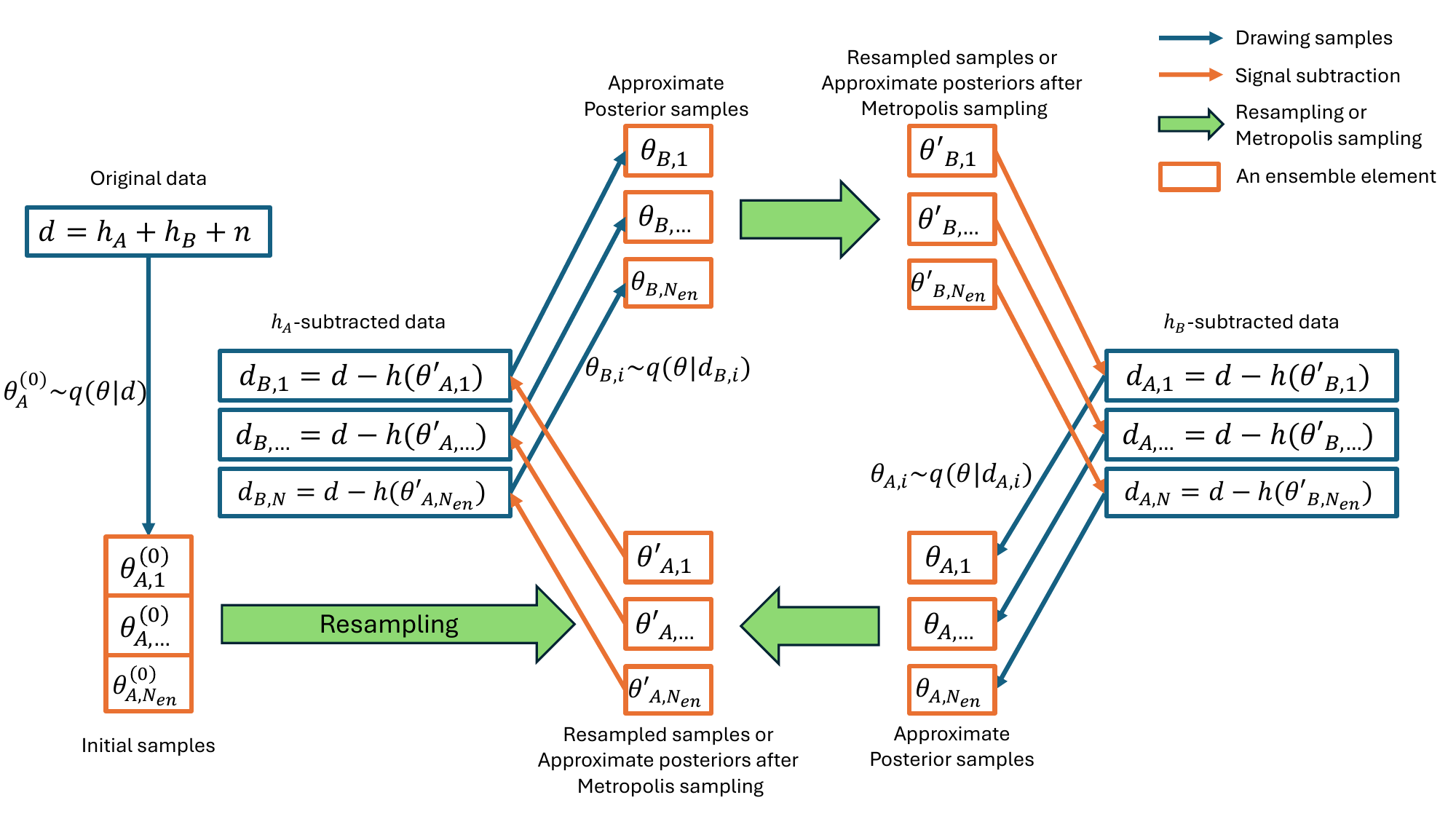}
  \caption{\label{fig:algorithm} Hierarchical subtraction workflow excluding the convergence part, more details are given in the main text and Algorithm \ref{alg}. The original data $d$ contains two signals $A$ and $B$, and we assume $A$ is first found by the NDE. Each initial sample is used to generate a signal for subtraction, and they evolve together as an ensemble in the hierarchical subtraction iteration. At first several iterations, resampling is used to remove low-likelihood elements in the ensemble. Then the resampling will be replaced by Metropolis sampling that leads the sampler to convergence. Concatenations of $\theta_{A, i}^{'}$ and $\theta_{B, i}^{'}$ for $i$ ranging from 1 to $N_\mathrm{en}$ are the final posterior samples for signal $A$ and $B$, respectively. }
\end{figure*}

The description of the hierarchical subtraction workflow is summarized as pseudocode in Algorithm \ref{alg} and (partly) visualized in Fig.~\ref{fig:algorithm}. 

\begin{algorithm}[H]
  \caption{Hierarchical subtraction for overlapping GW signals \label{alg}}
  \begin{algorithmic}
  
  \State \textbf{Input:} GW Data $d = h_A + h_B + n$, Neural Density Estimator $q$, Initial temperature $T = 10$, Max number of iterations $N_{\mathrm{it}} = 30$, Max number of retrial $N_\mathrm{trial}=100$, Convergence JSD threshold $\text{JSD}_\text{thre}=10^{-4}$ nat
  
  \State \textbf{Initialize:}
  \State Draw $\{ \theta_A^{(0)} \} \sim q(\theta \mid d)$ 
  \State Compute $\{ \log L^{(0)} \} = \max_{\{\phi_{c, A}^{(0)}\}} ( - \frac{1}{2} \| d - h(\{\theta_{A}^{(0)}\}) \|^2$ )
  \State Add phase $\{ \theta_A^{(0)} \} = \{ \theta_A^{(0)} \} \cup \{\phi_{c, A}^{(0)}\}_\mathrm{maxL}$
  \State $\{ \theta_A^{'(0)} \} = \text{Resampling}(\{ \theta_A^{(0)} \}, \{ \log L^{(0)} \}, T)$
  \State Compute $\bar{L}^{(0)} = \textrm{mean} (\{ \log L^{(0)} \})$
  
  \State $\text{DoResampling} = 1$, $\text{DoMetropolis} = 0$, 
  \State $X = A$, $Y = B$, $\{ \theta_X^{'(0)} \} = \{ \theta_A^{'(0)} \}$
  \State $n_{\mathrm{it}} = 1$
  \State \textbf{Hierarchical subtraction and sampling:}
  \While{$n_{\mathrm{it}} < N_{\mathrm{it}}$}
      \State Sample $\{ \theta_Y^{(n_{\mathrm{it}})} \} \sim q(\theta \mid d - h(\theta_X^{'(n_{\mathrm{it}}-1)}))$
          \State $\{ \log L^{(n_{\mathrm{it}})} \} =$
          \State $~~~~\max_{\{\phi_{c, Y}^{(n_{\mathrm{it}})}\}} \left( -\frac{1}{2} \| d - h(\{\theta_X^{(n_{\mathrm{it}}-1)}\}) - h(\{\theta_Y^{(n_{\mathrm{it}})}\}) \|^2 \right)$
          \State $\{ \theta_Y^{(n_{\mathrm{it}})} \} = \{ \theta_Y^{(n_{\mathrm{it}})} \} \cup \{\phi_{c, Y}^{(n_{\mathrm{it}})}\}_\mathrm{maxL}$
          \State $\bar{L}^{(n_{\mathrm{it}})} = \textrm{mean} (\{ \log L^{(n_{\mathrm{it}})} \})$
      \State $n_\mathrm{trial} = 0$
      \While{$\bar{L}^{(n_{\mathrm{it}})} < \bar{L}^{(n_{\mathrm{it}}-1)}$ and $\text{DoResampling}$}
          \State Repeat line 13-17
          \State $n_\mathrm{trial} = n_\mathrm{trial} + 1$
          \If{$n_\mathrm{trial} == N_\mathrm{trial}$}
              \State $\text{DoResampling} = 0$, $\text{DoMetropolis} = 1$
          \EndIf
      \EndWhile
  
      \If{$\text{DoResampling}$}
          \State $\{ \theta_Y^{'(n_{\mathrm{it}})} \} = \text{Resampling}(\{ \theta_Y^{(n_{\mathrm{it}})} \}, \{ \log L^{(n_{\mathrm{it}})} \}, T)$
          \State $T = \text{ExponentialAnnealing}(T)$ 
      \ElsIf{$\text{DoMetropolis}$}
          \State $(\{ \theta_Y^{'(n_{\mathrm{it}})}\}, \{\theta_X^{'(n_{\mathrm{it}}-1)} \}) = $
          \State \( \begin{aligned}
            \text{MetropolisSampling}(&\\
            &(\{ \theta_Y^{'(n_{\mathrm{it}})}\}, \{\theta_X^{'(n_{\mathrm{it}}-1)} \}), \\
            & (\{ \theta_Y^{'(n_{\mathrm{it}}-2)}\}, \{\theta_X^{'(n_{\mathrm{it}}-1)} \}),  \\
            &\{\log L^{(n_{\mathrm{it}})}\}, \\
            &\{\log L^{(n_{\mathrm{it}}-1)}\} \\
            &)
        \end{aligned} \)
      \EndIf
  
      \If{$\text{JSD}(\{ \theta_Y^{'(n_{\mathrm{it}})} \}, \{ \theta_Y^{'(n_{\mathrm{it}}-2)} \}) < \text{JSD}_\text{thre}$ \\ \hspace{1cm}and $\text{JSD}(\{ \theta_X^{'(n_{\mathrm{it}}-1)} \}, \{ \theta_X^{'(n_{\mathrm{it}}-3)} \}) < \text{JSD}_\text{thre}$ \\ \hspace{1cm}and $\text{DoMetropolis}$}
          \State $\theta_Y = \{ \theta_Y^{'(n_{\mathrm{it}})} \}$, $\theta_X = \{ \theta_X^{'(n_{\mathrm{it}}-1)} \}$
          \State \textbf{Output:} Hierarchical subtraction converges at iteration $n_{\mathrm{it}}$.
          \State \textbf{Exit}
      \Else
          \State Swap: $(X, Y) = (Y, X)$
          \State $n_{\mathrm{it}} = n_{\mathrm{it}} + 1$
      \EndIf      
  \EndWhile
  \State \textbf{Output:} Hierarchical subtraction does not converge within $N_{\mathrm{it}}$ iterations.
  \end{algorithmic}
\end{algorithm}

\bibliography{refs.bib}


\end{document}